\documentclass[aps,pra,reprint,superscriptaddress]{revtex4-1}
\usepackage{amsmath}
\usepackage{graphicx}
\usepackage{amsfonts}
\usepackage{color}
\usepackage{bm}
\usepackage{lipsum}
\usepackage{comment}

\usepackage{amsmath}
\usepackage{amssymb}
\usepackage{graphicx} 
\usepackage{braket}
\usepackage{mathtools}
\usepackage{times}
\usepackage{enumerate}

\newcommand{\ii}{\mathrm{i}}

\newcommand{\Tr}{\mathrm{Tr}}

\newcommand*{\defeq}{\mathrel{\vcenter{\baselineskip0.5ex\lineskiplimit0pt\hbox{\scriptsize.}\hbox{\scriptsize.}}}=}

\usepackage[colorlinks,linkcolor=blue,urlcolor=blue,citecolor=blue]{hyperref}

\begin{document}

\title{Multi-spin probes for thermometry in the strong-coupling regime}

\author{Marlon Brenes}
\email{marlon.brenes@utoronto.ca}
\affiliation{Department of Physics and Centre for Quantum Information and Quantum Control, University of Toronto, 60 Saint George St., Toronto, Ontario, M5S 1A7, Canada}

\author{Dvira Segal}
\affiliation{Department of Physics and Centre for Quantum Information and Quantum Control, University of Toronto, 60 Saint George St., Toronto, Ontario, M5S 1A7, Canada}
\affiliation{Department of Chemistry, University of Toronto, 80 Saint George St., Toronto, Ontario, M5S 3H6, Canada}

\date{\today}

\begin{abstract}
We study the sensitivity of thermometric probes that are composed of $N$ spins coupled to a sample prepared at temperature $T$. Our analysis extends beyond the weak-coupling limit into the strong sample-probe coupling regime. In particular, sample-induced interactions between each of the spins are generated via strong coupling effects and are not fine-tuned amongst each body composing the probe.  By employing the reaction-coordinate mapping to evaluate the non-canonical equilibrium state of the probe at finite coupling, we compute the thermometric sensitivity via the quantum Fisher information through the equilibrium state itself. We find that for single-spin probes $(N = 1)$, temperature sensitivity decreases in the regime of weak-to-intermediate coupling strength, however, as the coupling increases we observe much higher sensitivity of the probe in the low-temperature regime. Furthermore, as long as $N > 1$, there exist optimal values of the sample-probe interaction energy that allow one to attain enhanced thermometric sensitivity when compared to the maximum achieved precision obtained from thermal Gibbs states at weak coupling, particularly in the regime of low temperature. Finally, we show that this enhanced sensitivity may be observed from suboptimal measurements. 
\end{abstract}

\maketitle

\section{Introduction}
\label{sec:intro}

Temperature estimation in the quantum domain is a fervent research field, which has received theoretical and practical attention in recent years~\cite{Mehboudi:2019Rev,Giovannetti:2011Rev}. As a subset of the already growing field of quantum thermodynamics~\cite{Goold:2016Rev}, quantum thermometry has emerged to develop and understand precise protocols for temperature estimation at the nanoscale. Achieving high precision in the estimation of very low temperatures is a difficult task, with a number of applications ranging from cold-atomic systems for quantum simulation~\cite{Mitchison:2020Exp,Hofstetter:2018Rev} to sensing with nitrogen-vacancy centres in diamond~\cite{Fujiwara:2021Rev} and biological systems~\cite{Kucsko:2013Rev}.

Diverse directions have taken place as recourse to achieve high-precision thermometry, most of which fall into two categories: {\em local} and {\em global} thermometry. While {\em global} thermometry~\cite{Rubio:2021Global,Mok2021Global} rose as means to understand temperature estimation in situations where the temperature range is not well-known {\em a priori}, {\em local} thermometry concerns the design of temperature probes and the optimal measurements to be carried out to achieve high-temperature sensitivity~\cite{Mehboudi:2019Rev,Giovannetti:2011Rev,Correa:2015Individual,Mehboudi:2015Precision,Guo:2015Therm,DePasquale:2016Local,Campbell:2017Therm,Marcin:2018Fermion,Sone:2018Discord,Campbell:2018Therm,Mehboudi:2019Polaron,Hovhannisyan:2019MB,Sone:2019NonC,Latune:2020,Potts:2020Bound,Montenegro:2020Opto,Hovhannisyan:2021Optimal,Khan:2022Bose,Mihailescu2023Impurity,Abiuso:2023Optimal,Sone:2023Cond}. Adaptative-Bayesian strategies have also come into place with promising precision enhancement in temperature estimation~\cite{Boeyens:2021Bayesian,Mehboudi:2022Bayesian,Glatthard:2022Bayesian,Brask:2022Bayesian}. In non-integrable quantum systems, where thermalisation is ubiquitous, the eigenstate thermalisation hypothesis provides the means to estimate temperatures from local operations~\cite{Mitchison:2022ETH,Brenes:2020Fisher,Deutsch:1991,Srednicki:1994,Srednicki:1996,Srednicki:1999}. Quantum thermal machines have also been proposed as means for temperature estimation~\cite{Hofer:2017QTM,Levy:2020Machine}.

In turn, {\em local} thermometry can also be sub-categorised into two different classes of protocols: those which achieve temperature estimation via the study of the resulting equilibrium state from coupling a probe to a sample~\cite{Mehboudi:2019Rev} and those which do so via the out-of-equilibrium dynamical response signals~\cite{Johnson:2016Noneq,Mancino:2017Noneq,Kiilerich:2018Dyn,Mukherjee:2019,Cavina:2018Noneq,Feyles:2019Noneq,Mancino:2020Noneq}. We shall refer to the former as {\em equilibrium thermometry}, where temperature estimation may only follow from indirect measurements on the equilibrium state. The precision of the temperature estimation, in this case, will depend on both the equilibrium state itself and on the particular indirect measurement chosen. 

Whenever the probe-sample interaction energy is the smallest energy scale in the configuration, quantum master equations~\cite{BreuerPetruccione,Landi:2022Rev} predict that the equilibrium state of the probe, i.e., the resulting state in the limit of long times starting from a product state between a probe and a thermalised sample, will be a thermal Gibbs state. We refer to the Gibbs state as the ``canonical" state.
Certain microscopic conditions need to be met for the equilibrium state to be thermal~\cite{Michel:2007QME,Archak:2022QME,Archak:2023Lindblad}, although, thermalisation between a probe and a sample at weak interaction energy is a physical phenomenon that occurs with a high degree of universality. In the coupling regime where the equilibrium state of the probe is canonical, several aspects have been highlighted in order to employ these states for temperature estimation. The optimal measurement that provides the highest temperature sensitivity is the energy measurement of the probe~\cite{Mehboudi:2019Rev}, while the design of the probe that provides the ultimate temperature sensitivity is one for which the $M$ levels of the energy spectrum of the probe contains a single, non-degenerate ground state; together with a $(M - 1)$-degenerate excited state~\cite{Correa:2015Individual}. For practical purposes, achieving such a high degree of control and design is indeed very complicated. 

In the regime where the probe-sample interaction energy cannot be neglected, the equilibrium state of the probe is non-canonical~\cite{Cresser:2021Ultrastrong,Trushechkin:2022MFGS,Segal:2023Effective,Chiu:2022MFGS} and it has been discussed that energy measurements remain optimal even in this regime from perturbation theory~\cite{Miller:2023Pert}; while bath-induced correlations and strong coupling may lead to enhanced temperature sensitivity in integrable and harmonic models~\cite{Correa:2017Strong,Planella:2022Multi,Salado:2021Critical}. It has also been shown that non-Markovian effects, which may be prominent in the strong probe-sample interaction energy, could also lead to enhancements in temperature estimations from dynamical signals~\cite{Zhang:2021NM,Zhang:2022NM}.

In this work, we consider thermal probes that are composed of multiple spins and possibly strongly coupled to a sample as means for temperature estimation. In particular, we consider sample-induced spin interactions to determine whether an enhancement in the temperature estimation may be achieved. While the equilibrium state in the ultra-strong coupling regime may be accessed via the projection of the probe Hamiltonian onto the eigenbasis of the coupling operator between the sample and the probe~\cite{Cresser:2021Ultrastrong,Segal:2023Effective}; in the intermediate (non-perturbative) coupling regime, the equilibrium state is most appropriately described via numerical approaches. The reaction-coordinate mapping~\cite{Iles:2014RC,Strasberg:2016RC,Nick:2021RC,Hughes:2009RC,Hughes:2009RC2} may be employed in certain operational regimes with a high degree of accuracy~\cite{Iles:2014RC} for specific spectral functions of the sample~\cite{Martinazzo:2011RC} to compute the equilibrium state at strong coupling. The reaction-coordinate mapping provides the means to address strong-coupling effects via a Markovian embedding~\cite{Woods:2014Embedding}, in which an enlarged system Hamiltonian evolves under Markovian dynamics. It can also be extended via polaron transformations that allow for analytical insight~\cite{Segal:2023Effective}. We consider multi-spin probes coupled to a reaction coordinate to model strong-coupling effects and bath-mediated interactions. With this method, we study the reduced state obtained when tracing out the reaction-coordinate degrees of freedom, leading to a non-canonical equilibrium state of the probe. 

To address the temperature sensitivity, we consider the signal-to-noise ratio (SNR) as a figure of merit, which can be upper-bounded with the quantum Fisher information~\cite{Mehboudi:2019Rev,Pezze:2018Rev,Braunstein1994,Tth2014} through the quantum Cramer-Rao bound~\cite{Pezze:2018Rev}. By computing maximal SNR via the non-canonical equilibrium state of the probe mediated by bath-induced interactions (Fig.~\ref{fig:1}), we observe the following:
\begin{enumerate}[(i)]
    \item{For a single-spin probe, the effect of strong coupling is detrimental to the optimal temperature sensitivity of the probe at weak-to-intermediate coupling energy. This falls in line with the findings in Ref.~\cite{Miller:2023Pert} and extends the results therein to the non-perturbative regime of strong coupling. In the intermediate-to-strong coupling regime, much higher sensitivity may be observed in the low-temperature regime (Fig. \ref{fig:2}).}
    \item{For multi-spin probes where the internal interactions amongst each of the $N$ spins are mediated via bath-induced correlations, we find that, as long as $N > 1$, the temperature range over which the probe is sensitive increases considerably. Furthermore, there exists an optimal coupling $\lambda$ between the sample and the probe for a given temperature range and probe size $N$ to attain the optimal SNR 
    (Figs. \ref{fig:3}-\ref{fig:4}).}
    \item{The broad-range temperature sensitivity for multi-spin probes can be attained via dephasing operations (diagonal measurements) on the reduced state of the probe, at the cost of decreased sensitivity in the high-temperature regime, but not in the low-temperature regime. Local operations, such as polarisation measurements on the multi-spin probe, diminish temperature sensitivity in the low-temperature regime but the observed sensitivity is higher than the one obtained from energy measurements at weak coupling (Fig. \ref{fig:5}).}
\end{enumerate}
These results contribute to the growing literature to establish ultimate thermometric precision bounds in strong-coupling thermodynamics. Two drawbacks of equilibrium thermometry that have been pointed out are the long timescales required for equilibration~\cite{Correa:2015Individual} and the highly-peaked sensitivity at the level of the SNR that it is often observed~\cite{Mehboudi:2022Bayesian,Alves:2022Bayesian,Glatthard:2022Bayesian}, which leaves one to somehow obtain prior knowledge of the temperature range to be estimated. We argue that bath-mediated interactions may alleviate these constraints, by increasing the temperature range over which the probe provides high sensitivity and in many cases reducing the timescales of equilibration via strong-coupling dynamics.

In Sec.~\ref{sec:bounds} we introduce the common language of equilibrium thermometry and our reaction-coordinate mapping, as well as the model we employ for equilibrium thermometry. In Sec.~\ref{sec:probes} we delve into the optimal SNR results for our probe configurations and the SNR obtained from suboptimal measurements. We provide analysis and conclusions in Sec.~\ref{sec:conc}, together with some proposals for future directions. 

\section{Equilibrium thermometry}
\label{sec:bounds}

\subsection{Ultimate precision bounds}

Focusing on equilibration processes, thermometry relies on the parameter estimation from the equilibrium state of a {\em probe}. A thermalised sample is coupled to a probe and the entire configuration is allowed to relax to equilibrium. At weak coupling, the equilibrium state of the probe $(k_{\rm{B}} \defeq 1)$
\begin{align}
\label{eq:gibss_state}
\hat{\rho}_p(\beta) = \frac{e^{-\beta \hat{H}_p}}{Z_p},
\end{align}
depends on the parameter under investigation, in this case, the temperature $T \defeq 1 / \beta$. The temperature may only be estimated via a set of $m$ indirect measurements on the equilibrium state of the probe. The equilibrium state is defined via the Hamiltonian of the probe $\hat{H}_p$ and $Z_p(\beta) = \Tr[{{\rm exp}(-\beta \hat{H}_p)}]$.

The ultimate precision that may be attained via this parameter estimation protocol is understood from the Cramer-Rao inequality~\cite{Mehboudi:2019Rev}
\begin{align}
\label{eq:bound_precision}
\delta T \geq [m \mathcal{F}(T)]^{-1/2},
\end{align}
where $\delta T$ stands for the temperature precision and $\mathcal{F}$ is the temperature-dependent quantum Fisher information (QFI) which, in this context, may be understood as the {\em sensitivity} of each {\em optimal} measurement~\cite{Pezze:2018Rev}. The QFI is obtained when maximising the classical Fisher information (FI) over all possible measurements~\cite{Giovannetti:2011Rev}. Note that Eq.~\eqref{eq:bound_precision} is valid for any coupling strength.

It has been shown that the observable $\hat{O}$ with the largest (optimal) temperature sensitivity at thermal equilibrium is the Hamiltonian $\hat{H}_p$ of the probe itself, such that the minimum statistical uncertainty on the signal-to-noise ratio (SNR) is given by~\cite{Correa:2015Individual}
\begin{align}
(T / \delta T)^2 \leq m C(T),
\end{align}
where $C(T) = (\delta \hat{H}_p / T)^2$ is the heat capacity of the system and $\delta^2 \hat{H}_p = \Tr[\hat{\rho}_p(T)\hat{H}^2_p] - \Tr[\hat{\rho}_p(T)\hat{H}_p]^2$. We remark that the SNR can be related to the heat capacity of the probe only in the context of thermal equilibrium states. Whenever, the equilibrium state is non-canonical, the SNR cannot; in general, be related to the heat capacity of the probe~\cite{Mehboudi:2019Rev,Miller:2023Pert}.

In a more general sense, it will be more useful for our discussion to consider the QFI as~\cite{Pezze:2018Rev,Mehboudi:2019Rev,Giovannetti:2011Rev}
\begin{align}
\label{eq:qfi_l}
\mathcal{F}(\beta) = \Tr[\hat{L}_{\beta}^2 \hat{\rho}_p(\beta)],
\end{align}
where $\hat{L}_{\beta}$ is the symmetric-logarithmic derivative (SLD) defined implicitly from the Lyapunov equation 
\begin{align}
\label{eq:sld}
\partial_{\beta} \hat{\rho}_p(\beta) = \frac{1}{2} \{ \hat{L}_{\beta}, \hat{\rho}_p(\beta) \},
\end{align}
with $\{ \cdot, \cdot \}$ denoting the anti-commutator and $\partial_{\beta} \defeq \partial / \partial \beta$. Following our previous discussion, the most informative measurements can be shown to be the projections onto the eigenbasis of $\hat{L}_{\beta}$~\cite{Mehboudi:2019Rev}. For thermal equilibrium processes, where the equilibrium state is of the form $\hat{\rho}_p(\beta) = {\rm exp}(-\beta \hat{H}_p) / Z_p(\beta)$, the SLD can be shown to be $\hat{L}_{\beta} = \langle \hat{H}_p \rangle - \hat{H}_p$~\cite{Mehboudi:2019Rev}. In this case, the SLD is diagonal in the energy eigenbasis of the Hamiltonian of the probe. In Eq.~\eqref{eq:sld}, we have defined the SLD as a function of $\beta$. We note that as a function of temperature, the SLD obtains the form $\hat{L}_{T} = (\hat{H}_p - \langle \hat{H}_p \rangle) / T^2$.

If the equilibrium state of the probe is non-canonical, as may be the case when the sample-probe interaction energy is non-negligible, these results are not satisfied, in general~\cite{Miller:2023Pert}.

\begin{figure}[t]
\fontsize{13}{10}\selectfont 
\centering
\includegraphics[width=0.8\columnwidth]{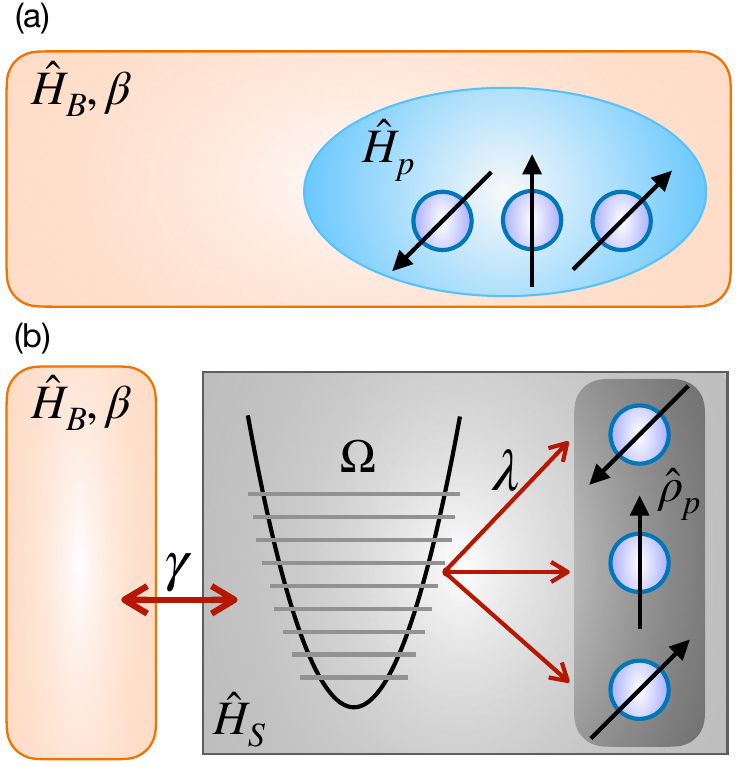}
\caption{(a) A few-qubit probe $\hat{H}_p$ is coupled to a thermal reservoir (sample) $\hat{H}_B$, of which the temperature $T = 1 / \beta$ is to be estimated. (b) A harmonic mode with frequency $\Omega$ is extracted from the reservoir and included as part of the probe with coupling $\lambda$. The residual coupling $\gamma$ between the enlarged probe $\hat{H}_S$ and $\hat{H}_B$ is a perturbative parameter in the configuration. This reaction coordinate mapping is an example of a {\em Markovian embedding}. We study the sensitivity of the probe from the reduced state after tracing out the reaction-coordinate degrees of freedom $\hat{\rho}_p(\beta) = \Tr_{\rm{RC}}[e^{-\beta \hat{H}_S}] / Z_S(\beta)$.}
\label{fig:1}
\end{figure}

\subsection{Thermalisation and strong-coupling thermal fixed point through the reaction coordinate mapping}

At weak coupling, the state of the probe $\hat{\rho}_p(\beta)$ may be seen as the steady state of the resulting dynamics between the probe to a sample, modelled as a thermal reservoir, of which the temperature is to be estimated. 
The total Hamiltonian of the configuration is given by
\begin{align}
\hat{H}_{\rm tot} = \hat{H}_p + \hat{H}_{\tt B} + \gamma \hat{H}_{\rm int},
\end{align}
where $\hat{H}_p$, $\hat{H}_{\tt B}$ and $\hat{H}_{\rm int}$ are the Hamiltonians of the probe, bath (sample) and their interaction; respectively. The coupling between the probe and the bath is controlled via the dimensionless parameter $\gamma$. In standard open-systems theory, a perturbative approximation to the second order of $\gamma$ together with the Born-Markov approximations yield a quantum master equation in Lindblad form for the dynamics of the probe~\cite{BreuerPetruccione}
\begin{align}
\label{eq:master_equation}
\partial \hat{\rho}_p / \partial t = -\ii[\hat{H}_p, \hat{\rho}_p] + \mathcal{L}\{ \hat{\rho}_p \},
\end{align}
where $\mathcal{L}$ is the Lindblad superoperator and $[\cdot, \cdot]$ is the commutator. The above equation dictates the effective dynamics of the probe from environmental effects. For a given physical configuration, the form of $\mathcal{L}$ will depend on the microscopic details of the probe-bath interaction Hamiltonian and a thermodynamically-consistent treatment has to be considered for Eq.~\eqref{eq:master_equation} to yield the correct steady-state at long-times, i.e., the (canonical) thermal state in Eq.~\eqref{eq:gibss_state}~\cite{Landi:2022Rev,Michel:2007QME,Archak:2022QME}. Most importantly, the approximations that lead to Eq.~\eqref{eq:master_equation} require the probe and the bath to approximately be in a product state throughout the dynamics and that correlation functions of the bath decay over timescales much shorter than the characteristic timescales of the dynamics of the probe~\cite{BreuerPetruccione}. 

At strong coupling, beyond second-order perturbative approximations in $\gamma$, these conditions cannot be guaranteed~\cite{Cresser:2021Ultrastrong}. Instead, in this regime, the {\em total} equilibrium state is a Gibbs state of the entire configuration
\begin{align}
\hat{\rho}_{\rm{tot}}(\beta) = \frac{e^{-\beta \hat{H}_{\rm{tot}}}}{Z_{\rm{tot}}(\beta)},
\end{align}
such that the {\em reduced} state of the probe is the partial trace over environmental degrees of freedom~\cite{Cresser:2021Ultrastrong}
\begin{align}
\hat{\rho}_p(\beta) = \frac{\Tr_{\tt{B}}\left[e^{-\beta \hat{H}_{\rm{tot}}}\right]}{Z_{\rm{tot}}(\beta)},
\end{align}
the complication being that this expression leaves one to describe the, in principle, infinite amount of degrees of freedom of the environment. In certain scenarios, however, one may instead consider the repartitioning of the Hamiltonian into an enlarged system that contains certain degrees of freedom of the bath, and a residual bath to which the system is coupled. The mapping becomes useful as long as the resulting enlarged Hamiltonian remains weakly coupled to the residual bath. This approach is typically known as a {\em Markovian embedding}~\cite{Woods:2014Embedding}, whereby strong-coupling effects are captured via the explicit evolution of the probe's state combined with some bath degrees of freedom. An example of a specific type of Markovian embedding is the so-called reaction coordinate mapping, as depicted in Fig.~\ref{fig:1}~\cite{Iles:2014RC,Strasberg:2016RC,Hughes:2009RC,Hughes:2009RC2,Nick:2021RC}.

Consider a probe coupled to a bosonic bath modelled via an infinite set of harmonic oscillators with total Hamiltonian
\begin{align}
\label{eq:original_model}
\hat{H}_{\rm tot} = \hat{H}_p + \hat{S} \sum_k f_k (\hat{b}^{\dagger}_k + \hat{b}_k) + \sum_k \nu_k \hat{b}^{\dagger}_k \hat{b}_k,
\end{align}
where the set of $\{ \hat{b}_k \}$ are canonical bosonic operators for the $k$-th mode with frequency $\nu_k$ and $f_k$ is the coupling strength between probe $\hat{H}_p$ and sample through the probe's operator $\hat{S}$. The reaction-coordinate mapping in its most basic form starts by extracting a collective mode (with canonical bosonic operators $\{ \hat{a} \}$) from the bath and including it as part of the system, such that the probe is turned into an enlarged system
\begin{align}
\label{eq:mapping}
\hat{H}_p + \Omega \hat{a}^{\dagger} \hat{a} + \lambda \hat{S} (\hat{a}^{\dagger} + \hat{a}) \mapsto \hat{H}_S,
\end{align}
where the extended system $\hat{H}_S$ is now weakly-coupled to the {\em residual} bath, i.e., the resulting bath description after the extraction of the strongly-coupled mode. In Eq.~\eqref{eq:mapping}, $\lambda$ is the coupling strength and $\Omega$ the frequency of the extracted mode. Both $\lambda$ and $\Omega$ follow from the spectral function of the original (previous to the mapping) sample $J(\omega)$, via~\cite{Strasberg:2016RC,Segal:2023Effective}
\begin{align}
\label{eq:lambda_omega}
\lambda^2 = \frac{1}{\Omega} \int_0^\infty {\rm{d}}\omega \; \omega J(\omega), \quad \Omega^2 = \frac{\int_0^\infty {\rm{d}}\omega \; \omega^3 J(\omega)}{\int_0^\infty {\rm{d}}\omega \; \omega J(\omega)}.
\end{align}
The mapping can be shown to lead to an extended system coupled weakly to a residual bath for certain spectral densities of the original model Eq.~\eqref{eq:original_model}~\cite{Strasberg:2016RC}. If that is the case, then one can justify a master equation of the form Eq.~\eqref{eq:master_equation} that leads to appropriate thermalisation of the \emph{extended system}, such that in the limit of long times the steady state is thermal $\hat{\rho}_S(\beta) = {\rm exp}(-\beta \hat{H}_S) / Z_S(\beta)$, where $Z_S(\beta) = \Tr[{{\rm exp}(-\beta \hat{H}_S)}]$ and $\hat{H}_S$ is the Hamiltonian of the enlarged system.

\begin{figure}[t]
\fontsize{13}{10}\selectfont 
\centering
\includegraphics[width=1\columnwidth]{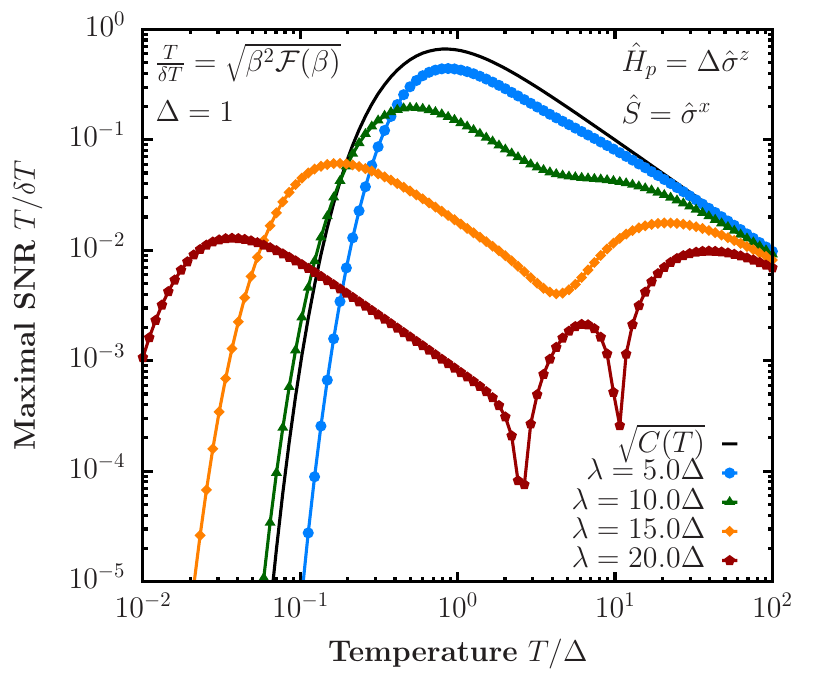}
\caption{Maximal signal-to-noise ratio (SNR) for a single-spin probe as a function of temperature. The solid black line is the weak coupling ultimate precision bound given by the heat capacity of the probe, while the rest of the curves depict the SNR at strong-coupling parameters $\lambda = 5\Delta, 10\Delta, 15\Delta, 20\Delta$. The reaction coordinate with frequency $\Omega = 15\Delta$ is truncated to $M = 50$ levels which is sufficient to guarantee convergence.}
\label{fig:2}
\end{figure}

\begin{figure*}[t]
\fontsize{13}{10}\selectfont 
\centering
\includegraphics[width=1\textwidth]{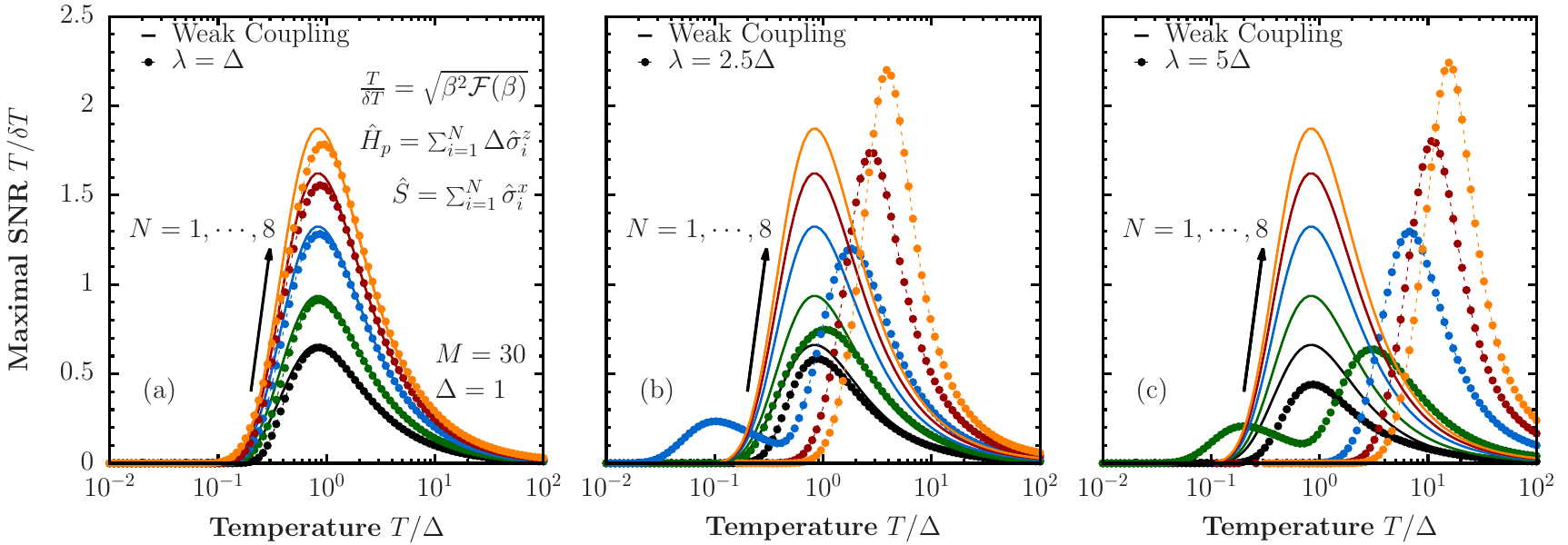}
\caption{Maximal signal-to-noise ratio (SNR) for multi-spin probes as a function of temperature. The reaction coordinate with a frequency of $\Omega = 15\Delta$ is truncated to $M = 30$ levels such that the curves have converged in the temperature range selected. Panels (a), (b), and (c) show the SNR for different coupling strengths $\lambda = \Delta, 2.5\Delta$, and $5\Delta$, respectively. The curves are shown for different values of $N = 1,2,4,6,8$, increasing from bottom to top as displayed in the panels. Solid lines correspond to weak-coupling ($\lambda\to 0$) results, while dashed-dotted lines depict the results in the strong-coupling regime for each value of $\lambda$.}
\label{fig:3}
\end{figure*}

Through this approach, one may investigate strong-coupling thermometric effects by studying the reduced state of the probe after tracing out the reaction-coordinate degrees of freedom
\begin{align}
\label{eq:eq_state_red}
\hat{\rho}_p(\beta) = \frac{\Tr_{\rm RC}[e^{-\beta \hat{H}_S}]}{Z_S(\beta)},
\end{align}
and computing first the SLD through Eq.~\eqref{eq:sld} and then QFI for the reduced state through Eq.~\eqref{eq:qfi_l}.

\section{Spin probes}
\label{sec:probes}
The Hamiltonian of the model is given by Eq. \eqref{eq:original_model}.
We consider a probe composed of $N$ spins with the Hamiltonian 
\begin{align}
\label{eq:ham_probe}
\hat{H}_{p} = \sum_{i=1}^{N} \Delta \hat{\sigma}^z_i.
\end{align}
The spins composing the probe do not interact with each other, however, they are coupled strongly to the bath with a system operator given by
\begin{align}
\label{eq:s_coupling}
\hat{S} = \sum_{i=1}^{N} \hat{\sigma}^x_i.
\end{align}
Using the reaction-coordinate mapping, we define the system Hamiltonian Eq.~\eqref{eq:mapping} including the original probe model,
the reaction coordinate, and their mutual interaction.
The reaction coordinate itself couples to the residual bath allowing thermalisation of the extended system. For details on the mapping see e.g., Ref.~\cite{Nick:2021RC}.

The extraction of a reaction coordinate from the bath (sample) and its inclusion as part of the probe (thermometer) Hamiltonian, as written in Eq. \eqref{eq:mapping} elucidates the generation of an effective coupling between all pairs of spins in the limit of non-vanishing coupling $\lambda$. This is the case since the spins in $\hat{H}_S$, which are otherwise non-interacting, are coupled via a collective operator $\hat S$ {\it to the same reaction-coordinate} mode. This degree of freedom, which is included explicitly in the equilibrium state in Eq.~\eqref{eq:eq_state_red} before being traced out, mediates couplings between the spins of the probe.

As mentioned before, the system composed of the probe and the reaction coordinate is assumed to thermalise to a canonical Gibbs state. The reduced state of the probe can be shown to thermalise to a Gibbs state at weak $\lambda$~\cite{Segal:2023Effective}, however, this is not necessarily the case as $\lambda$ increases. A natural question is thus whether strong coupling effects in our model could lead to enhanced or detrimental maximal signal-to-noise ratios, which we can compute via
\begin{align}
\frac{T}{\delta T} = \sqrt{\beta^2 \mathcal{F}(\beta)}
\end{align}
in the single-shot scenario $(m = 1)$.

\subsection{Single-spin probe}

The most elemental form of spin probes, corresponding to a single spin strongly coupled to the bath, serves as a basic benchmark. In this case, we consider $N = 1$ in Eq.~\eqref{eq:ham_probe}.
The extended system Hamiltonian includes a single spin coupled to a reaction-coordinate mode (the latter coupled to the sample). For the SNR, we consider the reduced state of the single spin induced by strong coupling. 

We compute the SNR as a function of temperature for different values of the coupling parameter $\lambda$. The results are shown in Fig.~\ref{fig:2}. The calculation involves the computation of the SLD $\hat{L}_{\beta}$ though Eq.~\eqref{eq:sld} to then compute the QFI in Eq.~\eqref{eq:qfi_l}, with $\hat{\rho}_p = \Tr_{\rm RC}[e^{-\beta \hat{H}_S}] / Z_S$ and $\hat{H}_S$ from Eq.~\eqref{eq:mapping}. The reaction coordinate with a frequency of $\Omega = 15\Delta$ is truncated to $M = 50$ levels, which was sufficient to attain convergence of the results shown in Fig.~\ref{fig:2}.
The value of the reaction coordinate frequency emerges from the characteristics of the spectral function of the bath (sample), see 
Eq. \eqref{eq:lambda_omega} and Appendix~\ref{ap:freq}.

At weak coupling, the maximal SNR for the single-spin probe can be shown to be related to the heat capacity of the probe through $\sqrt{C(T)}$~\cite{Correa:2015Individual}, as depicted in the solid black line in Fig~\ref{fig:2}. We have that $C(T) = \partial_T \langle \hat{H}_p \rangle$ and can be computed analytically to obtain the maximum SNR in the single-shot scenario
\begin{align}
\frac{T}{\delta T} = \sqrt{C(T)} = \frac{2\Delta \beta e^{\beta \Delta}}{1 + e^{2\beta \Delta}}.
\end{align}
We see in Fig.~\ref{fig:2}, that the effect of strong coupling is detrimental to the sensitivity of single-spin thermometers in the weak-to-intermediate coupling regime ($\lambda \lesssim 5\Delta$). This falls in line with the results exposed in Ref.~\cite{Miller:2023Pert}, in which a perturbative treatment lead to the conclusion that energy measurements in the weakly-coupled case remain the most informative measurements.

However, as the coupling strength $\lambda$ increases, we see that at low temperatures, stronger coupling in the single-spin probe leads to much higher sensitivity than its weakly-coupled counterpart. In fact, in the range $T / \Delta = [10^{-2}, 10^{-1}]$, strong coupling leads to a SNR several orders of magnitude higher than the heat capacity of the spin-probe at weak coupling. A fast decay is observed at a given temperature for all the curves, indicating that one can only achieve certain precision at given temperature ranges from this protocol.

Most interestingly, though, the reduced state of the probe $\hat{\rho}_p$ does not acquire off-diagonal elements in this model~\cite{Segal:2023Effective}, which means that both $\hat{\rho}_p(\beta)$ and $\hat{L}_{\beta}$ are diagonal operators. This indicates that the precision shown in Fig.~\ref{fig:2} can be achieved via the measurements of the populations of the spin-probe at strong-coupling and there exists no need to evaluate the optimal basis for the measurement, as it corresponds to simple occupations of the reduced density matrix of the probe at equilibrium.

\begin{figure*}[t]
\fontsize{13}{10}\selectfont 
\centering
\includegraphics[width=1\textwidth]{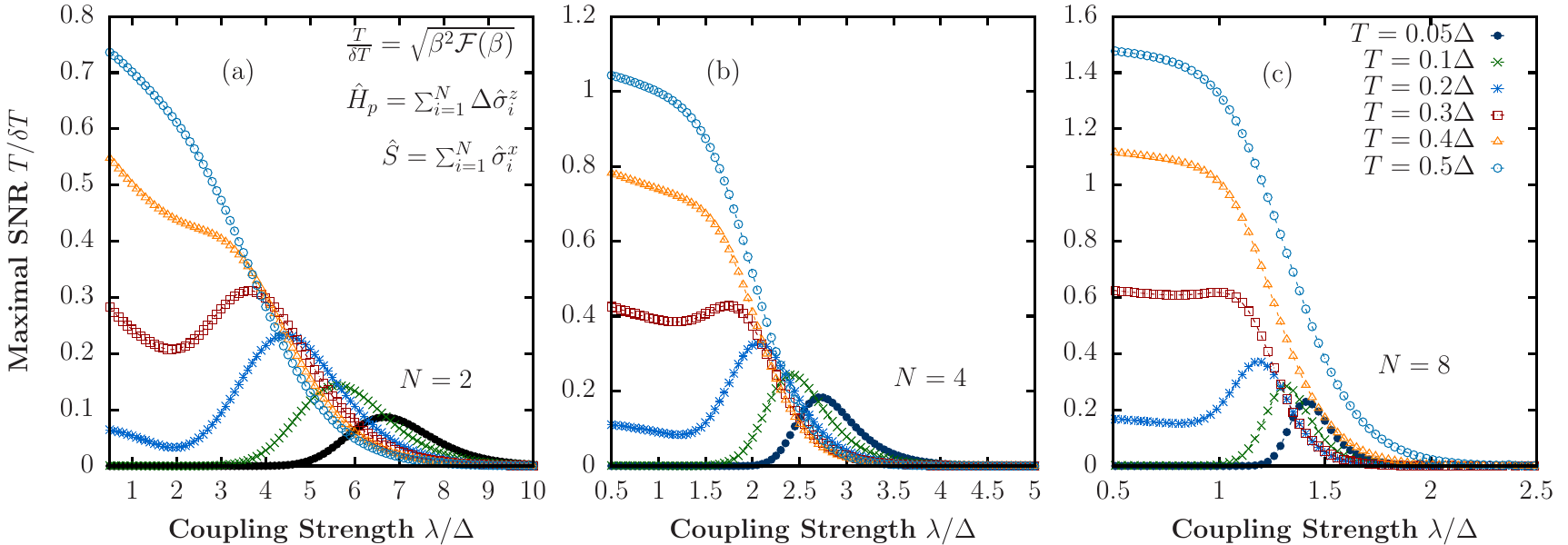}
\caption{Maximal signal-to-noise ratio (SNR) for multi-spin probes as a function of the coupling strength. The reaction coordinate with a frequency of $\Omega = 15\Delta$ is truncated to $M = 30$ levels such that the curves have converged in the selected range. Panels (a), (b), and (c) show the SNR for different numbers of spins $N = 2, 4$, and $8$, respectively.}
\label{fig:4}
\end{figure*}

We can gather from these results that at intermediate-to-strong coupling, the populations of the spin levels acquire a different temperature dependence than the canonical ones, translating to differences in the SLD compared to weak coupling. This distinct dependence translates to an increased sensitivity of the probe at a lower temperature for sufficiently strong $\lambda$. Interestingly, for our choice of $\hat{S} = \hat{\sigma}^x$, no coherences are generated in the reduced state of the probe. Different choices for the coupling operator $\hat{S}$ do indeed lead to temperature-dependent coherences in the state of the probe. The choice of the coupling operator %$\hat{S}$ 
can largely affect the sensitivity of the probe at strong coupling. See Appendix~\ref{ap:change_s_single} for further details. 

\subsection{Multi-spin probes}

Having understood the single-spin probe at strong coupling, we now turn our attention to the $N > 1$ case. Recalling Eq.~\eqref{eq:ham_probe} and Eq.~\eqref{eq:s_coupling}, we do not allow spins to directly interact with each other. However, they do develop an effective interaction via their strong coupling to the sample.

Figure \ref{fig:3} shows the SNR results as a function of temperature for different $N$ and different coupling parameters $\lambda$. It can be seen from Fig.~\ref{fig:3}(a) that at weak-intermediate coupling, the behaviour of the SNR is rather similar to the one observed for the single-probe case. The optimal measurements remain to be the energy measurements in the basis of the weakly-coupled probe, even in the multi-spin case. In fact, in this regime of weak coupling, the bath-mediated interactions are weak enough that each of the spins composing the probe barely interact with each other. The increased sensitivity follows the trivial $\sqrt{N}$ scaling for uncorrelated spins~\cite{Pezze:2018Rev} at $\lambda \to 0$, which can be confirmed from Fig.~\ref{fig:3}(a).

However, as shown in Fig.~\ref{fig:3}(b) and Fig.~\ref{fig:3}(c),  the effect of strong coupling is non-trivial. Particularly, the maximal SNR can be larger at strong coupling than its weakly-coupled counterpart, albeit at higher temperature ranges. Furthermore, strong coupling reveals different SNR peaks at different temperature ranges for certain values of $N$. We thus come to the conclusion from these results that temperature sensitivity may be higher at strong coupling for multi-spin probes, unlike the single-probe configuration. Furthermore, this effect can only be observed at relatively strong coupling as a collective effect stemming from many-body interactions induced by the sample.

In Fig.~\ref{fig:4} we show the SNR as a function of the coupling strength $\lambda$ for probes composed of a different number of spins at different temperature values. These results show that indeed, at low temperatures, strong coupling translates to a higher sensitivity in multi-probe configurations. In fact, strong coupling translates to temperature sensitivity in certain regimes where weakly-coupled probes provide negligible information through energy measurements. Furthermore, there exists an optimal coupling at a given temperature for which the sensitivity is maximal. This effect washes away as the temperature increases, where we recover that the most informative measurements are the ones related to weakly-coupled configurations. We then see that multi-spin probes are primed for low-temperature thermometry. 

\subsection{Suboptimal measurements}

\begin{figure*}[t]
\fontsize{13}{10}\selectfont 
\centering
\includegraphics[width=1\textwidth]{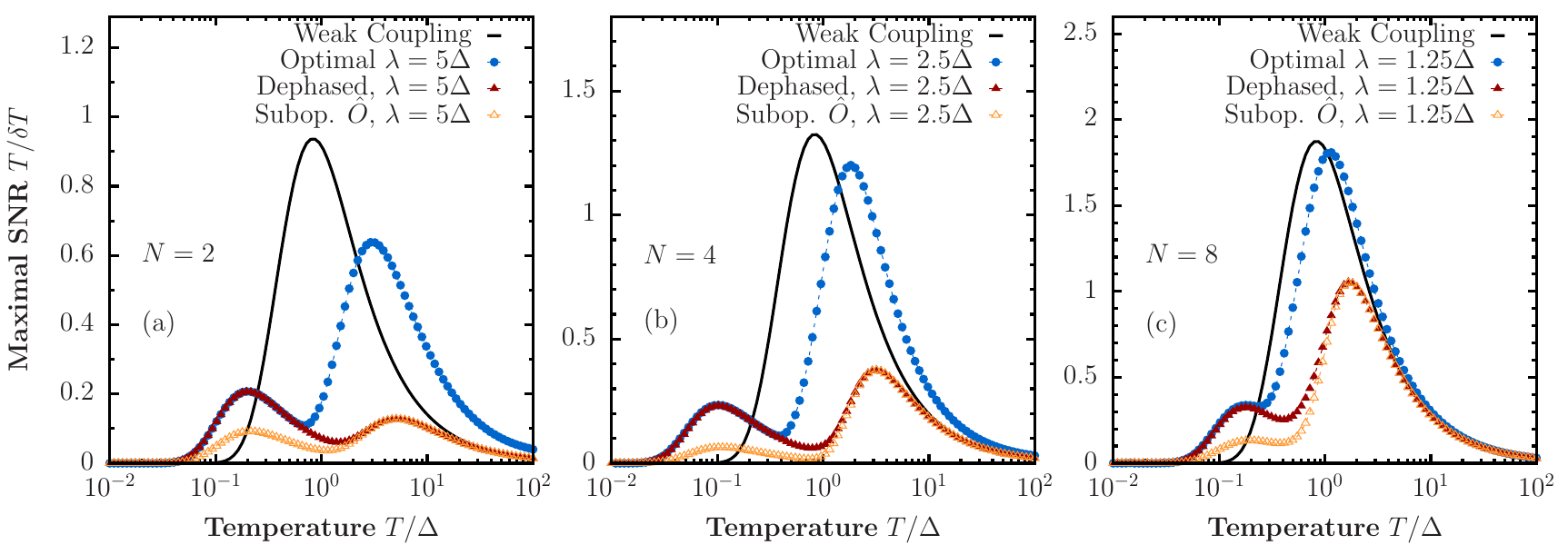}
\caption{Maximal signal-to-noise ratio (SNR) for multi-spin probes as a function of the temperature for different measurement schemes. The reaction coordinate with a frequency of $\Omega = 15\Delta$ is truncated to $M = 30$ levels such that the curves have converged in the temperature range selected. Panels (a), (b) and (c) show the SNR for different probes with different number of spins, $N = 2, 4$, and $8$, respectively. The values of the coupling $\lambda$ are chosen such that the strong-coupling SNR is larger than its weak-coupling ($\lambda\to 0$) counterpart (see Fig.~\ref{fig:4}).}
\label{fig:5}
\end{figure*}

We have seen that for a single spin probe, strong coupling increases thermometric sensitivity at sufficiently strong $\lambda$ in the low-temperature regime. On the other hand, multi-spin probes show interesting behaviour, whereby strong coupling and many-body effects may provide higher temperature sensitivity, particularly in the low-temperature regime. However, achieving such precision from the equilibrium states can be very complicated. Indeed, even at weak coupling, energy measurements can involve highly non-local operations which pose practical and technical complications. At strong coupling, to take advantage of the higher sensitivity that may exist in certain temperature ranges as we have seen in Figs.~\ref{fig:3} and~\ref{fig:4}, the situation is even more complicated. In this coupling regime, the SLD develops off-diagonal matrix elements which are temperature-dependent, the same as the equilibrium state $\hat{\rho}_p$ from Eq.~\eqref{eq:eq_state_red}. Therefore, it is even more difficult to understand and choose the optimal basis which renders the SLD in diagonal form, which then fixes the basis for the measurements required to attain the fundamental bound for thermometry. It is therefore imperative to consider suboptimal measurements, ones that are more feasible from the experimental perspective. 

In light of this, we now consider the temperature sensitivity of the spin probes from suboptimal measurements. The first operation we consider is the dephasing of the reduced state of the probe $\hat{\rho}_p$ onto its diagonal basis
\begin{align}
\tilde{\rho}_p = \sum_k \ket{k}\bra{k} \hat{\rho}_p \ket{k}\bra{k},
\end{align}
where $\ket{k} = (0,\cdots,1_k,\cdots,0)^T$ are basis vectors such that $\tilde{\rho}_p$ is diagonal in the spin basis. From this state we can compute the Fisher information via
\begin{align}
\label{eq:f_d}
\mathcal{F}_{\rm D}(\beta) = \Tr [\tilde{L}^2_{\beta} \tilde{\rho}_p(\beta)],
\end{align}
where $\tilde{L}_{\beta}$ is the SLD from Eq.~\eqref{eq:sld} computed through $\tilde{\rho}_p(\beta)$. Naturally, $\tilde{L}_{\beta}$ is also diagonal in the spin basis, which then implies that the optimal SNR $T / \delta T = \sqrt{\beta^2 \mathcal{F}_{\rm D}(\beta)}$ follows from the estimation of the occupations (diagonal matrix elements) of $\hat{\rho}_p(\beta)$.  

We may also consider the suboptimal measurements which follow from the estimation of a local observable $\hat{O}$. Given an observable $\hat{O}$, we may consider the SNR from the measurements following the expectation values of $\hat{O}$, via
\begin{align}
\label{eq:f_c}
\frac{T}{\delta T} = \frac{T |\chi_T(\hat{O})|}{\delta \hat{O}} \leq \sqrt{\beta^2 \mathcal{F}(\beta)}
\end{align}
in the single-shot scenario ($m = 1$). In Eq.~\eqref{eq:f_c}, $\delta^2 \hat{O} \defeq \langle \hat{O}^2 \rangle - \langle \hat{O} \rangle^2$ is the variance of $\hat{O}$ in the reduced state of the probe $\hat{\rho}_p(\beta)$ and $\chi_T(\hat{O}) \defeq \partial_{\alpha} \Tr[\hat{\rho}_p(\alpha) \hat{O}]|_{\alpha=T}$ is the temperature susceptibility of $\hat{O}$~\cite{Mehboudi:2019Polaron}. We choose an extensive, yet local operator, for the suboptimal measurement. Consider an extensive sum of the spin polarizations in the $z$ component,
\begin{align}
\label{eq:subop_o}
\hat{O} \defeq \sum_{k=1}^{N} \hat{\sigma}^z_k,
\end{align}
which amounts to estimating the total polarisation of the probe composed of $N$ spins.

In Fig.~\ref{fig:5} we display the maximal signal-to-noise ratio $T / \delta T$ as a function of temperature for different system sizes $N$ and couplings $\lambda$, using four different measurement schemes: energy measurements at weak coupling, the optimal measurement at strong coupling at the value of $\lambda$, the dephasing operation which amounts to estimating the occupations of the density matrix and a sum of local operations where one measures the polarisation of all the spins in the $z$ direction. Figures \ref{fig:5}(a), \ref{fig:5}(b) and \ref{fig:5}(c) show different system sizes, $N = 2, 4$, and $8$, respectively. We start by highlighting that energy measurements at weak coupling have a higher peak for larger system sizes while, as discussed before, strong coupling effects lead to a multi-peaked SNR as a function of $T$ in the optimal basis. Remarkably, even considering the dephasing operation in the spin basis of $\hat{\rho}_p(\beta)$ leads to the same multi-peaked behaviour, at the cost of reduced sensitivity at higher temperatures. Furthermore, this operation retains the low-temperature sensitivity observed from the optimal measurements at the value of $\lambda$, so one need not consider the optimal bound on the SNR at strong coupling for low-temperature thermometry. Considering an extensive sum of local observables, however,  indeed leads to decreased sensitivity in the low-temperature regime. We do note that even this operation from local measurements, leads to broader-temperature sensitivity, with higher values of the SNR than the optimal weakly-coupled counterpart.

Our results suggest that high-temperature sensitivity increases with the system size $N$. This shows that many-body effects and sample-induced correlations in this case lead to relatively high sensitivity even from conducting local operations. Finally, we note that the optimal coupling $\lambda$ changes with both the system size and the temperature range over which the probe is sensitive. We have selected the values of $\lambda$ in Fig.~\ref{fig:5} such that they lie close to the optimal value taken from Fig.~\ref{fig:4}.

\section{Conclusions}
\label{sec:conc}

We have studied the impact of strong-coupling effects on 
equilibrium thermometry employing multi-spin probes.
Using the reaction-coordinate mapping, 
we showed that a non-canonical equilibrium state of such probes stems from strong-coupling effects with the sample. While the reduced states of the probes are non-canonical, the equilibrium state of the extended Hamiltonian that contains the reaction coordinate is indeed a standard Gibbs state and therefore, we can take advantage of the Markovian embedding to analyse the reduced states of the probes. This approximation can be shown to yield the correct equilibrium states in certain regimes of the spectral function of the sample~\cite{Segal:2023Effective,Nick:2021RC,Iles:2014RC}. From this treatment, we can consider strong-coupling effects in the non-perturbative regime. We have shown that, along the lines of the findings in Ref.~\cite{Miller:2023Pert} in which a perturbative treatment was employed, weak-to-intermediate coupling leads to an equilibrium state for which the optimal SNR for temperature thermometry is lower than its weakly-coupled, thermal Gibbs state counterpart. At stronger values of the coupling parameter, however, thermometric sensitivity in the case $N = 1$ is higher in the low-temperature regime. We can conclude from these results that the most informative measurements for single-spin probes are the energy measurements of the states that undergo a thermalisation process, up until the value of the sample-probe coupling strength increases such that one may attain higher sensitivity at low $T$. 

For multi-spin probes and, in particular, configurations where the probe is composed of spins that are not finely tuned to interact with each other but rather through the sample itself, we have found that strong coupling leads to enhanced precision in the low-temperature regime. This trend is in accord with another configuration, where each body comprising the probe is a harmonic mode and the interaction amongst each mode with the sample is quadratic in bosonic operators, such that the equilibrium state is a non-canonical Gaussian state. As shown in Ref.~\cite{Planella:2022Multi}, in such a configuration, low-temperature sensitivity is also enhanced via sample-induced correlations between each body comprising the probe. 
In both harmonic and anharmonic (spin) cases, the enhanced temperature sensitivity may also be accessed via local measurements. These results suggest that, in a more general sense, bath-induced correlations between local probes enhance low-temperature thermometry.
Our study, however, demonstrates that in multi-spin setups the SNR depends on the number of spins in a highly non-monotonic manner once the interaction extends beyond weak coupling.
Furthermore, in our model, the coupling mechanism of the spins to the sample provides another route for tunability of thermometric sensitivity (see Appendix~\ref{ap:freq}).
%Unlike harmonic probes, a multi-spin thermometer has multiple 'knobs' for optimization of its SNR at different temperature regimes. This includes the number of spins, their coupling mechanism to the sample and its magnitude, aspects that (beyond weak coupling) should be selected  in accord with the spectral properties of the sample.

With respect to practical applications, we envision our setup to be relevant in experimental platforms such as nitrogen-vacancy centres in diamond. Negatively-charged nitrogen-vacancy centres may be realised in diamond layers that, individually, act as individual spins with a high degree of tunability with respect to their effective self-energy and their inter-spin interaction using microwave pulses~\cite{Kucsko:2013Rev}. Sample-mediated correlations may be prominent in certain regimes of operation in these platforms. The relevant energy scale to ascertain the regime of temperatures over which the probe is sensitive corresponds to the splitting $\Delta$ between down and up states of each of the spins. Typically, these splittings can be measured in the 1-4 GHz range~\cite{Hernandez:2022NV,Wolf:2023NV}. Considering that in our setup, the multi-spin probes remain sensitive in the temperature regime of $T \sim [10^{-1}\Delta, 10\Delta]$ (see, e.g., Fig.~\ref{fig:5}), the use of multi-spin probes in the strong-coupling regime could lead to enhanced temperature sensitivities in the range between 1 mK and 100 mK. This range could be enhanced to sub-mK thermometry via further tuning of the energy splitting of the spins. Furthermore, through luminescence measurements, these experiments allow for full-state tomography of reasonably-sized spin clusters. While optimal energy measurements are indeed complicated, our description of suboptimal measurements is important for these practical approaches. A benefit of using these platforms is related to the long-coherent times available, typically longer than 1 ms~\cite{Balasubramanian:2009NV}, while, at strong coupling, equilibration occurs over much shorter timescales.

A direction that has been less explored is related to the effects on the temperature sensitivity from strong coupling effects from dynamical signals. Non-Markovian effects may be prominent in this regime~\cite{Rivas:2010NM,Zhang:2022NM}. Furthermore, dynamical signals in the non-Markovian regime differ substantially from their weakly-coupled counterparts and may be studied in certain regimes with the reaction-coordinate mapping~\cite{Iles:2014RC}. Given that non-Markovian effects may lead to enhanced temperature sensitivity~\cite{Zhang:2021NM}, a promising direction could be to consider strong-coupling effects on thermometric probes from the dynamical perspective using reaction-coordinate mapping.   

\begin{acknowledgments}
We gratefully acknowledge fruitful discussions with John Goold and Mark T. Mitchison. The work of M.B.~has been supported by the Centre for Quantum 
Information and Quantum Control (CQIQC) at the University of Toronto. D.S.~acknowledges support from NSERC and from the Canada Research Chair program. Computations were performed on the Niagara supercomputer at the SciNet HPC Consortium. SciNet is funded by: the Canada Foundation for Innovation; the Government of Ontario; Ontario Research Fund - Research Excellence; and the University of Toronto. 
\end{acknowledgments}

\appendix

\section{Maximally-coherent $\hat{S}$ for the single-probe case}
\label{ap:change_s_single}

\begin{figure}[t]
\fontsize{13}{10}\selectfont 
\centering
\includegraphics[width=1\columnwidth]{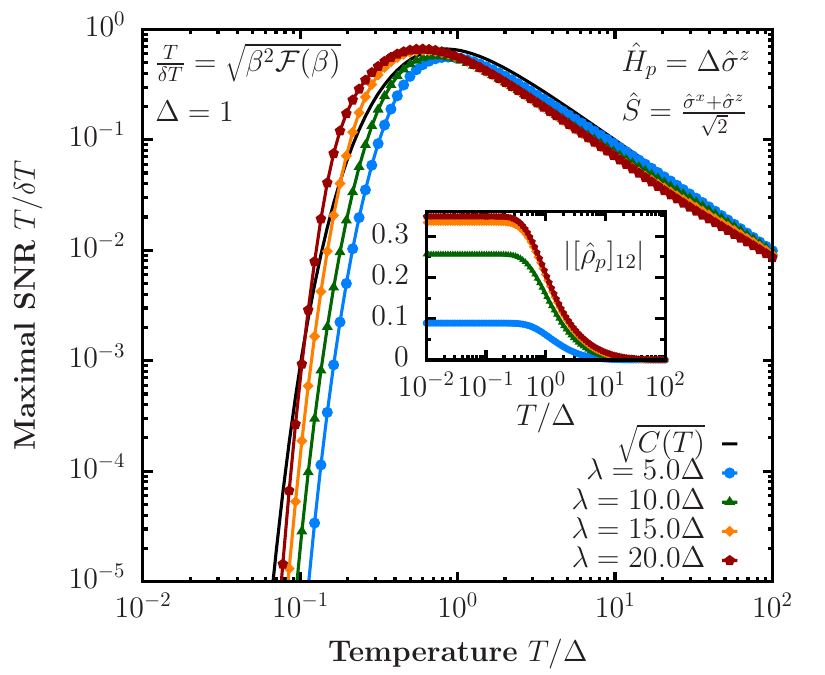}
\caption{Maximal signal-to-noise ratio (SNR) for a single-spin probe as a function of temperature. The solid black line is the weak coupling ultimate precision bound given by the heat capacity of the probe, while the rest of the curves depict the SNR at strong-coupling parameters $\lambda = 5\Delta, 10\Delta, 15\Delta, 20\Delta$. The reaction coordinate with frequency $\Omega = 15\Delta$ is truncated to $M = 50$ levels which is sufficient to guarantee convergence. Note the change of $\hat{S}$ with respect to Fig.~\ref{fig:2}. The inset shows the coherence of the reduced state of the probe $\hat{\rho}_p$ as a function of temperature for the same values of $\lambda$.}
\label{fig:ap1}
\end{figure}

In the main text, we have considered a specific type of probe-sample coupling interaction. From Eq.~\eqref{eq:original_model} and Eq.~\eqref{eq:s_coupling}, we have $\hat{S} = \hat{\sigma}^x$ for the single-spin probe. This choice is, in principle, arbitrary. One may instead consider different forms of probe coupling operators that lead to coherences in the spin basis of the reduced state of the single-spin probe $\hat{\rho}_p$ [Eq.~\eqref{eq:eq_state_red}]. For instance, if we consider
\begin{align}
\label{eq:s_coupling_coherent}
\hat{S} = \frac{1}{\sqrt{2}} ( \hat{\sigma}^x + \hat{\sigma}^z ),
\end{align}
then coherences develop in the spin basis of the reduced state of the probe $\hat{\rho}_p = \Tr_{\rm{RC}}[e^{-\beta \hat{H}_S}] / Z_S$ (see Eq.\eqref{eq:original_model}). In turn, these coherences become temperature-dependent~\cite{Segal:2023Effective} in the non-canonical equilibrium state of the probe at finite sample-probe interaction energy.

In Fig.~\ref{fig:ap1} we display the SNR for the spin-boson model (single-spin probe) as a function of temperature for different values of the coupling parameter $\lambda$. These results are analogous to the ones displayed in Fig.~\ref{fig:2} and we use the same parameters for the calculation, with the only difference being the coupling operator from Eq.~\eqref{eq:s_coupling_coherent}. It can be observed that for this type of coupling operator, the temperature sensitivity behaves quite differently at strong coupling than for the case $\hat{S} = \hat{\sigma}^x$ examined in the main text. In particular, the observed higher SNR at low temperatures disappears in this case. The inset in Fig.~\ref{fig:ap1} displays the coherences being developed in the spin basis of the reduced state of the probe, which vanish in the limit $T \to \infty$. In analogy to the results exposed in our main example in Fig.~\ref{fig:2}, at weak-to-intermediate coupling, it is the weakly-coupled Gibbs state SNR (solid black line in Fig.~\ref{fig:ap1}) that translates to the highest temperature sensitivity. However, as $\lambda$ increases, higher sensitivity may be achieved from the non-canonical equilibrium states of the probe. Nevertheless, we see that in this case, the SNR is not higher in the low-temperature regime when compared to its weakly-coupled counterpart, irrespective of the temperature-dependent coherences that develop in the reduced states of the probe $\hat{\rho}_p(\beta)$ at low temperature. We remark that the thermal Gibbs state at weak coupling is the equilibrium state of the probe irrespective of the sample-probe interaction operator, while the equilibrium state at strong coupling heavily depends on the microscopic details via $\hat{S}$~\cite{Cresser:2021Ultrastrong}. This implies that, at strong coupling, the microscopic details of the probe-sample interaction play an important role in the sensitivity of the probes. 

A recent work in Ref.~\cite{Ullah:2023Coherence} has found that coherences induced on the equilibrium state of a single-spin probe via interaction with ancillary spins may lead to enhancement in the SNR in the low-temperature regime at weak coupling. In this proposal, the ancillary spins are coupled to the sample but the probe is not. We remark that, in such configuration, the coherences that develop on the equilibrium state of the probe correspond to an effect of its interaction with the ancilla and are not generated through strong interaction with the sample. In such configuration, then, these coherences are different in nature than the ones we have study here, as they encode different processes. In Ref.~\cite{Ullah:2023Coherence}, low-temperature sensitivity was achieved by considering the energy splitting of ancillary spins to be much smaller than the energy spllitting of the probe. In such a way, these different energy scales set different parameters for temperature estimation. In these configurations, however, we expect the time to reach the equilibrium state to be rather long. Together with our results, we can further conclude that the microscopic details of the interaction between the probe and the sample are rather important towards high-precision thermometry.

\begin{figure}[t]
\fontsize{13}{10}\selectfont 
\centering
\includegraphics[width=1\columnwidth]{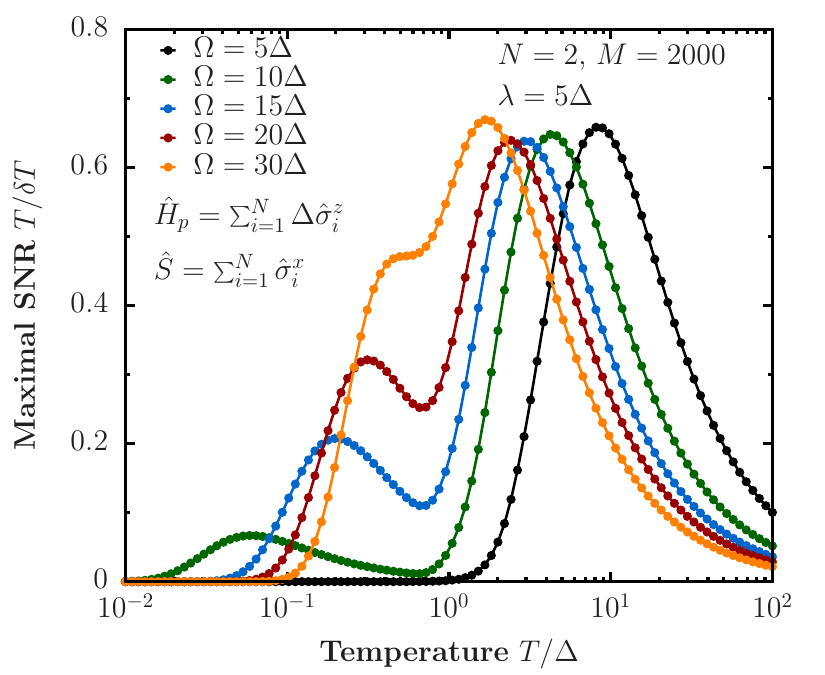}
\caption{Maximal signal-to-noise ratio (SNR) for the two-body $(N = 2)$ case as a function of temperature for different values of the frequency of the reaction coordinate $\Omega$. In this case, we employed $M = 2000$ which sufficed to guarantee convergence in the temperature regimes shown.}
\label{fig:ap2}
\end{figure}

\section{Effect of the spectral function of the sample}
\label{ap:freq}

An important parameter in our simulations is the natural frequency of the reaction coordinate, which we have denoted with $\Omega$ and described in Eq.~\eqref{eq:lambda_omega}. On physical grounds, $\Omega$ is the frequency of a collective-effective harmonic mode pertaining to the sample, to which the probe is most strongly coupled~\cite{Iles:2014RC}. Tuning this parameter yields different equilibrium states of the probe in the finite-$\lambda$ regime, as it is a property of the sample via its spectral function~\cite{Strasberg:2016RC}. A spectral density of Brownian form
\begin{align}
J(\omega) = \frac{4\gamma \Omega^2 \lambda^2 \omega}{(\omega^2 - \Omega^2)^2 + (2\pi\gamma\Omega\omega)^2},
\end{align}
which is peaked around $\Omega$ with width $\gamma$, leads to an effective spectral density, after the reaction-coordinate mapping, of the Ohmic type $J_{\rm{RC}}(\omega) = \gamma \omega e^{-|\omega| / \Lambda}$, where $\Lambda$ is a high-frequency cut-off~\cite{Iles:2014RC,Strasberg:2016RC,Nick:2021RC,Segal:2023Effective}. The dimensionless width parameter $\gamma$ is kept small, such that the enlarged system, comprising the probe and the reaction coordinate, is weakly coupled to the residual bath, i.e., to the sample after the reaction-coordinate mapping. In Fig.~\ref{fig:ap2} we display the SNR for the two-body spin probe $(N = 2)$ as a function of temperature for different values of $\Omega$. It can be observed that the effect of reducing $\Omega$ is to shift the temperature sensitivity to lower-temperature regimes, up to the point in which the low-temperature sensitivity vanishes for sufficiently low $\Omega$. In our calculations we kept the dimension of the manifold of the reaction coordinate to a very high value, $M = 2000$, to ensure convergence in the entire temperature regime shown in Fig.~\ref{fig:ap2}. From these results we can conclude that in employing multi-spin probes for temperature estimation in the strong-coupling regime, two important parameters are required to be considered hand-in-hand: the effective probe-sample coupling parameter $\lambda$ and the frequency of the collective harmonic mode of the sample to which the probe is most-strongly coupled. In a more general sense, this is the result of the equilibrium state at finite coupling depending strongly on the microscopic details of both the sample and its interaction with the probe, unlike thermal Gibbs states at weak coupling which are independent on these details.

\bibliography{bibliography}

\end{document}